\documentclass[conference,9pt]{IEEEtran}
\IEEEoverridecommandlockouts

\usepackage{cite}
\usepackage{amsmath,amssymb,amsfonts}
\usepackage{algorithmic}
\usepackage{graphicx}
\usepackage{textcomp}
\usepackage[table]{xcolor}

\usepackage{booktabs}
\usepackage[hidelinks]{hyperref}
\usepackage{siunitx}
\usepackage{mathtools}
\usepackage{comment}
\usepackage{multirow}
\usepackage{makecell}
\usepackage{arydshln}   
\usepackage{enumitem}

\makeatletter
\let\MYcaption\@makecaption
\makeatother
\usepackage[font=footnotesize,subrefformat=parens]{subcaption}
\makeatletter
\let\@makecaption\MYcaption
\makeatother

\usepackage{tikz}
\usetikzlibrary{patterns,patterns.meta}

\usepackage{xspace}
\makeatletter
\DeclareRobustCommand\onedot{\futurelet\@let@token\@onedot}
\def\@onedot{\ifx\@let@token.\else.\null\fi\xspace}

\makeatother

\def\equationautorefname~#1\null{(#1\null)}

\renewcommand{\sectionautorefname}{Section}
\renewcommand{\subsectionautorefname}{\sectionautorefname}

\let\orgautoref\autoref
\providecommand{\Autoref}[1]
{%
\def\figureautorefname{Figure}%
\def\subfigureautorefname{Figure}%
\orgautoref{#1}%
}
\renewcommand{\autoref}[1]
{%
\def\figureautorefname{Fig.}%
\def\subfigureautorefname{\figureautorefname}%
\def\sectionautorefname{Sec.}%
\def\subsectionautorefname{\sectionautorefname}%
\def\subsectionautorefname{\sectionautorefname}%
\orgautoref{#1}%
}

\def\appendixautorefname~#1\null{~#1 \null}

\makeatletter
\newcommand{\figcaption}[1]{\def\@captype{figure}\caption{#1}}
\newcommand{\tblcaption}[1]{\def\@captype{table}\caption{#1}}
\makeatother

\makeatletter 
\newcommand{\linebreakand}{%
  \end{@IEEEauthorhalign}
  \hfill\mbox{}\par
  \mbox{}\hfill\begin{@IEEEauthorhalign}
}
\makeatother 

\def\BibTeX{{\rm B\kern-.05em{\sc i\kern-.025em b}\kern-.08em
    T\kern-.1667em\lower.7ex\hbox{E}\kern-.125emX}}

\definecolor{E4002B}{HTML}{E4002B}
\usepackage{xparse}
\usepackage{contour}
\contourlength{0.1pt}
\NewDocumentCommand{\hatch}{O{E4002B} O{0.6pt} m}{%
  \begingroup
    \edef\HatchColor{#1}%
    \contourlength{#2}%
    \tikz[baseline=(txt.base),inner sep=0pt,outer sep=0pt]{%
      \node[
        pattern={Hatch[line width=0.8pt, distance=2pt, angle=45]},
        pattern color=E4002B,
        inner sep=2pt, outer sep=0pt,
        opacity=0.4,
      ] {\phantom{#3}};
      \node[inner sep=0pt, outer sep=0pt] (txt)
            {\contour{white}{#3}};
    }%
  \endgroup
}

\newcommand{\loosedata}{\textsc{Compound-mixed}\xspace}
\newcommand{\tightdata}{\textsc{Compound-tight}\xspace}
    
\begin{document}
\abovedisplayskip=1pt

\setlength\textfloatsep{8pt}
\setlength\dbltextfloatsep{8pt}
\setlength\floatsep{8pt}
\setlength\dblfloatsep{8pt}
\setlength\abovecaptionskip{2pt}
\setlength\belowcaptionskip{2pt}
\captionsetup[subfloat]{skip=0pt,aboveskip=2pt,belowskip=3pt}

\title{Can We Really Repurpose Multi-Speaker ASR Corpus\\for Speaker Diarization?}
\author{\IEEEauthorblockN{Shota Horiguchi, Naohiro Tawara, Takanori Ashihara, Atsushi Ando, and Marc Delcroix}
\IEEEauthorblockA{NTT, Inc., Japan}}

\maketitle

\begin{abstract}
Neural speaker diarization is widely used for overlap-aware speaker diarization, but it requires large multi-speaker datasets for training.
To meet this data requirement, large datasets are often constructed by combining multiple corpora, including those originally designed for multi-speaker automatic speech recognition (ASR).
However, ASR datasets often feature loosely defined segment boundaries that do not align with the stricter conventions of diarization benchmarks.
In this work, we show that such boundary looseness significantly impacts the diarization error rate, reducing evaluation reliability.
We also reveal that models trained on data with varying boundary precision tend to learn dataset-specific looseness, leading to poor generalization across out-of-domain datasets.
Training with standardized tight boundaries via forced alignment improves not only diarization performance, especially in streaming scenarios, but also ASR performance when combined with simple post-processing.
\end{abstract}


\section{Introduction}
Speaker diarization is the task of identifying speaker-wise speech activities in an input audio recording.
It can serve as prior knowledge for downstream tasks.
For example, in guided source separation, speaker-wise speech activities are used as constraints when estimating time-frequency masks to separate each speaker's utterances~\cite{boeddeker2018front,raj2023gpu}.
In multi-speaker automatic speech recognition (ASR), some models directly generate speaker-specific transcriptions using audio and the corresponding diarization results~\cite{polok2024but,polok2026dicow}.
More recently, speaker diarization has also been leveraged to generate training data for full-duplex spoken dialogue systems~\cite{defossez2024moshi}.

Recent approaches to speaker diarization commonly involve feeding the mixture audio directly into a neural network to obtain speaker-wise speech activities~\cite{fujita2019end1,medennikov2020targetspeaker,kinoshita2021integrating,kinoshita2021advances,horiguchi2022encoderdecoder,wang2023target,harkonen2024eend}.
Training such models requires datasets annotated with speaker-specific speech activities.
However, due to the limited availability of annotated data, early studies relied on simulated data for training~\cite{fujita2019end1}.
As the naturalness of the simulated data significantly affects model performance, various improvements have been proposed to simulation techniques~\cite{yamashita2022improving,landini2022from,landini2023multi}.
Nevertheless, the domain gap between simulated and real data remains a challenge.
Meanwhile, with the recent increase in publicly available multi-speaker audio data with annotations, an alternative approach has emerged: combining multiple datasets into a large training corpus~\cite{bredin2023pyannote,plaquet2023powerset,han2025leveraging}.
This compound-data approach provides a straightforward solution by eliminating the need for quality-sensitive simulation data.
In this study, we focus on this promising strategy.

\begin{figure}[t]
\centering
\includegraphics[width=\linewidth]{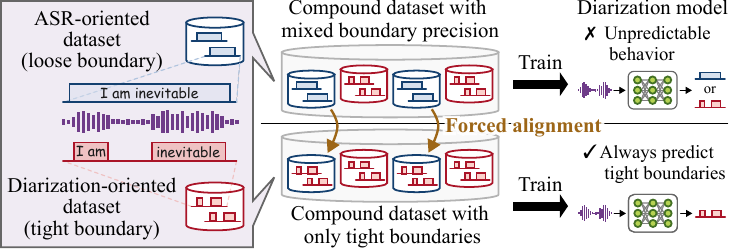}
\caption{Conventional training pipeline using compound dataset with mixed boundary precision (upper) and pipeline with only tight boundaries (lower).}
\label{fig:overview}
\end{figure}

Datasets that comprise compound data are not necessarily constructed solely for the purpose of training or evaluating speaker diarization systems.
It is common to repurpose corpora originally developed, at least in part, for multi-speaker ASR~\cite{bredin2023pyannote,plaquet2023powerset,han2025leveraging}.
While datasets specifically designed for diarization provide strictly annotated segment boundaries, corpora for multi-speaker ASR often prioritize semantic coherence, frequently merging utterances even when there are relatively long pauses.
When evaluating diarization systems on such repurposed corpora, the system may be penalized for accurately detecting pauses that were intentionally ignored in the original annotation.
As a result, the diarization error rate (DER) can worsen precisely because the model performs well in detecting speaker turns.
This makes meaningful evaluation of diarization performance difficult, yet the issue has been largely overlooked in prior studies~\cite{bredin2023pyannote,plaquet2023powerset,han2025leveraging,horiguchi2025pretraining}.

Moreover, models trained on such data may exhibit undesirable characteristics.
Since even long pauses must be treated as part of a speech segment, the model is forced to rely on longer contextual information to determine whether the pause is actually a part of a speech segment.
This is not preferable for, e.g., streaming inference, which requires prompt decisions based on incrementally arriving input.
In addition, when datasets with different labeling standards are combined for training, the model's inference behavior becomes unpredictable, posing practical challenges in real-world deployment.

In this paper, we conduct an extensive study on the impact of using loose boundaries in the evaluation of speaker diarization, as well as the effect of mixing boundary definitions when training diarization models with compound datasets.
\Autoref{fig:overview} summarizes the scope of our investigation.
First, we clarify that many pauses in ASR-oriented datasets are labeled as speech intervals by resegmenting them with forced alignment to create tighter boundaries.
This fact has been overlooked in prior work, but it represents a critical problem in which even ideal diarization can exhibit a DER of over \SI{20}{\percent}.
We then experimentally demonstrate that such labels are too loose for diarization and negatively affect model training, resulting in dataset-specific memorization of labeling precision.
Furthermore, we show that refining boundaries through forced alignment improves diarization performance in streaming inference.
Although strict segmentation may raise concerns about degrading ASR performance due to overly fragmented inputs, we also demonstrate that simple morphological closing~\cite{horiguchi2020utterance,boeddeker2023multi,boeddeker2024tssep} can effectively mitigate this issue and even enhance performance.
The labels obtained through forced alignment and used in this paper are available at \url{https://github.com/nttcslab-sp/diar-forced-alignment}.

\section{Related work}
\subsection{Speaker diarization datasets}
\label{sec:related_work_forced_alignment}
Real-world datasets used for evaluating speaker diarization systems can generally be categorized into two types based on their annotation methodology.
One type consists of datasets specifically developed for diarization purposes, such as DIHARD~\cite{ryant2019second,ryant2021third}, VoxConverse~\cite{chung2020spot}, and MSDWild~\cite{liu2022msdwild}.
We refer to these as \textit{DIA-oriented datasets} in this paper.
These datasets follow clearly defined annotation guidelines regarding segment boundaries, including criteria for determining the minimum pause length required to split a segment, as shown in \autoref{tbl:annotation}.

\begin{table}[t]
    \centering
    \setlength{\tabcolsep}{4pt}
    \caption{Annotation guideline of datasets used in diarization studies.}
    \label{tbl:annotation}
    \resizebox{\linewidth}{!}{%
    \begin{tabular}{@{}lll@{}}
        \toprule
        Name & Boundary annotation & Segment split strategy \\\midrule
        \multicolumn{3}{@{}l}{\textbf{Diarization-oriented datasets (Tight boundary)}}\\
        \:\:DIHARD III~\cite{ryant2021third}& Manual ($<$ \SI{10}{\ms} error) & Pauses $>$ \SI{200}{\ms}\\
        &\:\:or forced alignment\\
        \:\:VoxConverse~\cite{chung2020spot}& Manual ($<$ \SI{100}{\ms} error) & Pauses $>$ \SI{250}{\ms}\\
        \:\:MSDWild~\cite{liu2022msdwild}& Manual ($<$ \SI{100}{\ms} error) & Pauses $>$ \SI{250}{\ms}\\
        \:\:CHiME-6~\cite{watanabe2020chime} & Forced alignment & Pauses $>$ \SI{300}{\ms}\\\midrule
        \multicolumn{3}{@{}l}{\textbf{ASR-oriented datasets (Loose boundary)}}\\
        \:\:AISHELL-4~\cite{fu2021aishell} & Manual (No error bound) & Sentence\\
        \:\:AMI~\cite{carletta2007unleashing} & Manual (\qtyrange[range-units=single,range-phrase=--]{250}{500}{\ms} silence & Sentence\\
        & \:\:padding at both ends) \\
        \:\:AliMeeting~\cite{yu2022m2met} & Manual (No error bound) & Sentence\\
        \:\:CHiME-5~\cite{barker2018fifth} & Manual (No error bound) & Sentence\\
        \:\:DiPCo~\cite{van2020dipco}& Manual (No error bound) & Sentence (up to \SI{15}{\second})\\
        \:\:NOTSOFAR-1~\cite{vinnikov2024notsofar}&Manual (ASR-assisted)& ASR-based\\
        \bottomrule
    \end{tabular}%
    }
\end{table}

On the other hand, datasets developed for multi-speaker ASR, or \textit{ASR-oriented datasets}, often provide speaker-attributed speech segments along with corresponding transcriptions~\cite{fu2021aishell,carletta2007unleashing,yu2022m2met,barker2018fifth,van2020dipco,vinnikov2024notsofar}.
Because of this, they are frequently repurposed for research in speaker diarization~\cite{horiguchi2022encoderdecoder,landini2022bayesian,bredin2023pyannote,plaquet2023powerset}.
However, in contrast to DIA-oriented datasets, the segmentation criteria in ASR-oriented datasets are generally less strict.
They provide no guarantee that the boundary annotations fall within a certain error margin of the true segment boundaries.
Rather than aiming for temporal accuracy, segment boundaries are typically defined to maintain semantic continuity, avoiding splits that disrupt meaningful phrases or sentences.
For example, in the AMI corpus, annotators are instructed to insert short silence margins at segment boundaries and to avoid splitting segments except at natural linguistic boundaries such as sentence or phrase ends~\cite{moore2005guidelines}.
This design choice can be interpreted as part of an effort to facilitate smooth and coherent transcription.

\subsection{Speech activity annotation based on forced alignment}
In voice activity detection (VAD), a previous study compared manually annotated labels with those obtained through forced alignment based on ground truth transcriptions~\cite{kraljevski2015comparison}.
The study reported that the labels derived from forced alignment were more accurate than those produced by most human annotators, and their accuracy was comparable to that of expert annotators.
Based on this finding, using forced-alignment-based labels as ground truth has become a common practice in VAD research~\cite{tan2020rvad,bovbjerg2024self}.

In terms of speaker diarization, some discussions arose in the CHiME challenges, a series of multi-speaker ASR competitions.
CHiME-5 was held as a competition to evaluate transcription accuracy under the condition that speech segments were predefined based on human annotations~\cite{barker2018fifth}.
In contrast, CHiME-6, using the same recordings, required systems to jointly estimate both transcription and speech segmentation~\cite{watanabe2020chime}.
This update necessitated more precise segmentation to enable proper evaluation of speaker diarization.
Because the human annotations in the CHiME-5 data lacked a unified policy for segment boundary placement, the CHiME-6 data was constructed by applying forced alignment and splitting segments at pauses longer than \SI{300}{\ms}.
One CHiME-7 system paper reported the DER gap between reference labels with forced alignment and those without it~\cite{deng2023university}, but no discussion related to the gap was provided.
This study also applies forced alignment to correct ASR-oriented labels, but our contribution extends beyond this preprocessing step.
Specifically, i) we examine more extensively how the lack of resegmentation renders current diarization evaluations on ASR-oriented data fundamentally uninformative, and ii) we analyze more deeply how misaligned labels can adversely affect model training.

\section{Methodology}
This paper presents a comprehensive investigation into how ASR-oriented and DIA-oriented labels affect the training and evaluation of speaker diarization.
We use the same diarization pipeline based on end-to-end neural diarization with vector clustering (EEND-VC)~\cite{kinoshita2021integrating,kinoshita2021advances}, training it on different versions of compound data that differ only in annotation style.
In this section, we describe the forced alignment procedure employed to convert ASR-oriented labels into DIA-oriented ones, and the morphological closing operation used as post-processing to merge fragmented segments.

\subsection{Forced alignment for refinement of segment boundaries}
\begin{figure}[t]
\centering
\includegraphics[width=0.9\linewidth]{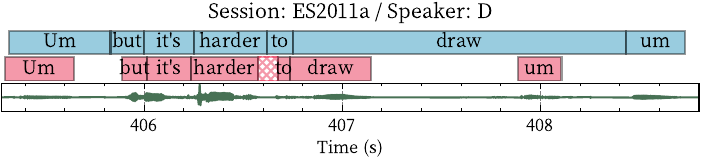}\\
\includegraphics[width=0.9\linewidth]{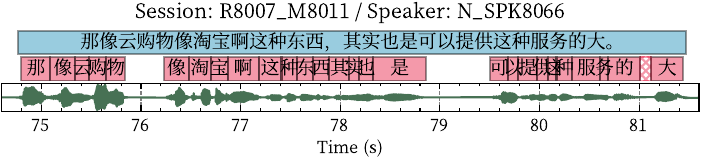}
\caption{Example of original segment (blue) and forced alignment (red) from AMI (top) and AliMeeting (bottom). Pauses shorter than \SI{200}{\ms}, which are treated as speech intervals under DIA-oriented labeling, are \hatch{hatched}.}
\label{fig:label_sample}
\end{figure}

Transcriptions typically accompany ASR-oriented datasets; therefore, it is reasonable to expect that the segment boundaries are subject to some level of quality control, but short pauses within segments are typically not captured.
When headset microphone recordings are available, forced alignment can be used to obtain word- or character-level annotations, following the standard practice in VAD research, as described in \autoref{sec:related_work_forced_alignment}.
This enables the resegmentation of each segment to eliminate short pauses that are typically ignored in the original annotations.
In this study, we leverage a publicly available pretrained model to relabel speech segments, thereby transforming ASR-oriented labels into ones that are more consistent with DIA-oriented annotation criteria.
Specifically, we apply this relabeling process to the AMI and AliMeeting corpora, which have headset recordings available and are commonly used for training and evaluation in EEND-VC-based approaches~\cite{bredin2023pyannote,plaquet2023powerset,han2025leveraging,horiguchi2025pretraining}.
Examples of the original ASR-oriented segments and forced-aligned segments in AMI and AliMeeting are shown in \autoref{fig:label_sample}.

The AMI corpus~\cite{carletta2007unleashing} is a collection of English conversations involving three to five speakers per session.
In this study, we use two types of recordings: one obtained by mixing the headset microphone recordings of all speakers (AMI-MHM) and the other taken from the single distant microphone recordings from the first channel of the microphone array (AMI-SDM).
While the corpus officially provides word-level timestamps obtained via forced alignment using the HTK toolkit~\cite{young1994htk}, it appears that pauses between words are not explicitly represented, as shown in \autoref{fig:label_sample}.
To obtain more precise segment boundaries, we employed Montreal Forced Aligner~\cite{mcauliffe2017montreal}, which is based on a GMM-HMM acoustic model, due to its superior alignment ability compared to other methods~\cite{rousso2024tradition}.
We used the pretrained English acoustic model~\cite{mfa_english_mfa_acoustic_2024} and dictionary~\cite{mfa_english_us_mfa_dictionary_2024} for alignment.
Note that pronunciations for out-of-vocabulary words in the transcript were generated using the grapheme-to-phoneme model~\cite{mfa_english_us_mfa_g2p_2023} in advance.

The AliMeeting corpus~\cite{yu2022m2met} is composed of Mandarin conversations of two to five speakers.
Since only the start and end times of each segment are provided, we compute character-level alignment using the Montreal Forced Aligner, with the pretrained acoustic model~\cite{mfa_mandarin_mfa_acoustic_2024} and dictionary~\cite{mfa_mandarin_china_mfa_dictionary_2024}.
In this case, too, the pronunciations of out-of-vocabulary words were estimated beforehand using the grapheme-to-phoneme model~\cite{mfa_mandarin_china_mfa_g2p_2024}.

After alignments were obtained, each pair of segments separated by short pauses less than \SI{200}{\ms} was merged, following the annotation guidelines of existing DIA-oriented datasets listed in \autoref{tbl:annotation}.

\subsection{Morphological closing for merging over-segmentation}\label{sec:closing}
From the perspective of ASR, over-segmentation of speech intervals can harm recognition performance.
Therefore, even if such segmentation is not optimal from a diarization perspective, one might argue that short pauses should be ignored and adjacent speech intervals merged to facilitate more robust ASR.
However, even when using DIA-oriented labels or inference results, it is possible to approximate ASR-oriented output by simply filling in short pauses.
In this study, we adopt morphological closing to treat short segments as part of continuous speech.
The closing operation involves extending boundaries of each speech segment by $m$ seconds on both sides, followed by shrinking both sides by $n$ seconds.
As a result, any pause shorter than $2m$ seconds is merged into the surrounding speech.
In our experiments, we set $m=n$ for simplicity.
This technique has already been adopted in several transcription systems as a practical post-processing step~\cite{horiguchi2020utterance,boeddeker2023multi,boeddeker2024tssep}.

\begin{figure}[t]
\centering
\includegraphics[width=\linewidth]{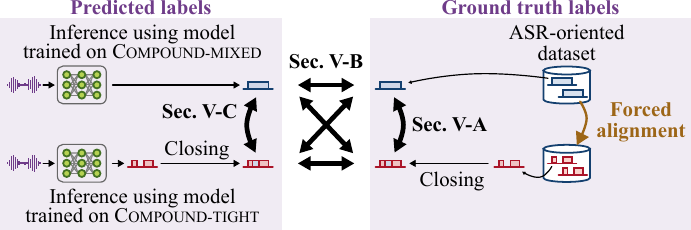}
\caption{Analyses performed using morphological closing.}
\label{fig:closing_patterns}
\end{figure}

In this study, we utilize closing for the following three types of analysis, each of which is also illustrated in \autoref{fig:closing_patterns}:
\begin{itemize}
\item We investigate whether applying closing to DIA-oriented reference labels---obtained via forced alignment from ASR-oriented data---can reconstruct the original ASR-oriented labels (\autoref{sec:result_forced_alignment}).
\item We evaluate whether applying closing to the output of a model trained on DIA-oriented labels can achieve DERs comparable to those obtained by a model trained on ASR-oriented labels, using ground truth as the reference (\autoref{sec:result_training}).
\item We assess dataset-wise pause-filling behavior by computing the closing width $\delta$ that minimizes the DER between the output of a model trained on mixed boundary precision and that of a model trained on tight boundaries with post-processing via closing (\autoref{sec:dynamics}).
\end{itemize}

\section{Experimental setup}
\subsection{Dataset}
For training and evaluation, we used two compound datasets, each composed of five domains as summarized in \autoref{tbl:dataset}.
For AMI-MHM, AMI-SDM, and AliMeeting, both the original ASR-oriented segments and the labels obtained via forced alignment were used.
In addition to these datasets, we included the following two DIA-oriented datasets for both compound sets: MSDWild~\cite{liu2022msdwild} and VoxConverse~\cite{chung2020spot}.
As a result, the \loosedata set uses the original labels for every component dataset, similar those used in the previous studies~\cite{bredin2023pyannote,plaquet2023powerset,han2025leveraging,horiguchi2025pretraining}.
This results in mixed precision of the segment boundaries among the component datasets, i.e., some samples have ASR-oriented labels, while others have DIA-oriented labels.
For \tightdata, on the other hand, each ASR-oriented dataset was relabeled by using forced alignment to ensure tight segment boundaries.
The total duration of the training set is about 356 hours for each.

\begin{table}[t]
\caption{Compound datasets used for model training.}\label{tbl:dataset}
\centering
\begin{tabular}{@{}lll@{}}
\toprule
Compound dataset&Component&Label\\\midrule
\loosedata&AMI-MHM &Original (ASR-oriented) \\
&AMI-SDM & Original (ASR-oriented)\\
&AliMeeting&Original (ASR-oriented)\\
&MSDWild (few)&Original (DIA-oriented)\\
&VoxConverse&Original (DIA-oriented)\\\midrule
\tightdata&AMI-MHM&Forced-aligned\\
&AMI-SDM&Forced-aligned\\
&AliMeeting&Forced-aligned\\
&MSDWild (few)&Original (DIA-oriented)\\
&VoxConverse&Original (DIA-oriented)\\
\bottomrule
\end{tabular}
\end{table}

For out-of-domain evaluation, we used AISHELL-4~\cite{fu2021aishell}, DiPCo~\cite{van2020dipco}, and NOTSOFAR-1~\cite{vinnikov2024notsofar} as ASR-oriented datasets, and DIHARD-3~\cite{ryant2021third} as a DIA-oriented dataset.

\subsection{Diarization pipeline}
Our diarization pipelines were built upon EEND-VC~\cite{kinoshita2021integrating,kinoshita2018listening}, implemented in the pyannote.audio framework~\cite{bredin2023pyannote}.
In the pipelines, local overlap-aware diarization was performed using a sliding window of 10 seconds in width and a 1-second shift, followed by speaker embedding clustering.

The goal of this study is not necessarily to achieve state-of-the-art performance, but rather to investigate how different datasets affect model training and evaluation.
Note that this conclusion remains valid because any system will be affected by the training labels, even ``perfect'' systems, as discussed in \autoref{sec:result_forced_alignment}.
To ensure the generality of our findings, we used two local diarization models with different levels of accuracy: a weaker SincNet followed by a 4-layer bidirectional long short-term memory (BLSTM)~\cite{bredin2023pyannote}, and a stronger ReDimNet-B2 with a single BLSTM~\cite{yakovlev2024reshape,horiguchi2025pretraining}.
Each model was trained to output frame-wise speech activities of four speakers with at most two-speaker overlap using the power-set loss~\cite{plaquet2023powerset}, resulting in a classifier with 11 output classes ($=\sum_{i=0}^2\binom{4}{i}$) per frame.
In this study, we did not adopt pretrained models based on self-supervised learning~\cite{han2025leveraging} or speaker identification~\cite{horiguchi2025pretraining}, as our focus is on investigating the impact of label quality in the training data on model performance.

For speaker embedding, we used an ECAPA-TDNN-based extractor~\cite{desplanques2020ecapatdnn} trained on the VoxCeleb 1\&2 datasets~\cite{nagrani2020voxceleb}.
The extracted embeddings were clustered using agglomerative hierarchical clustering with centroid linkage.
The clustering hyperparameters, i.e., clustering threshold and minimum cluster size, were tuned using a compound development set and shared across datasets during inference.

\section{Results and discussions}
\subsection{Impact of forced alignment on segment boundaries}\label{sec:result_forced_alignment}
\begin{table}
\centering
\caption{Discrepancy between ASR-oriented labels and those based on forced alignment, measured by DER and its breakdown, with the former as reference and the latter as prediction. MI: missed detection, FA: false alarm, CF: speaker confusion.}
\label{tbl:label_difference}
\begin{tabular}{@{}llcccc@{}}
\toprule
Dataset&Subset& MI & FA & CF & DER\\\midrule
AMI&Train&23.17\tiny{$\pm$3.65}&1.48\tiny{$\pm$0.71}&0.32\tiny{$\pm$0.19}&24.97\tiny{$\pm$3.75}\\
&Dev&19.05\tiny{$\pm$4.15}&1.83\tiny{$\pm$1.27}&0.29\tiny{$\pm$0.25}&21.18\tiny{$\pm$4.84}\\
&Test&22.88\tiny{$\pm$6.27}&1.36\tiny{$\pm$0.62}&0.36\tiny{$\pm$0.27}&24.60\tiny{$\pm$6.07}\\\midrule
AliMeeting&Train&13.77\tiny{$\pm$2.54}&0.00\tiny{$\pm$0.00}&0.00\tiny{$\pm$0.00}&13.77\tiny{$\pm$2.54}\\
&Dev&13.09\tiny{$\pm$1.86}&0.00\tiny{$\pm$0.00}&0.00\tiny{$\pm$0.00}&13.09\tiny{$\pm$1.86}\\
&Test&14.36\tiny{$\pm$2.31}&0.00\tiny{$\pm$0.00}&0.00\tiny{$\pm$0.00}&14.36\tiny{$\pm$2.31}\\
\bottomrule
\end{tabular}
\end{table}

We first assessed the difference between ASR-oriented labels, i.e., the ground truth used in previous studies, and those generated using forced alignment.
While both labels can be considered correct from the perspectives of ASR and diarization, respectively, we aimed to highlight their differences.
To this end, we calculated DERs using the transcription-based labels as the ground truth and the forced-alignment-based labels as predictions.
This setup clarifies how the current evaluation framework would score a diarization system that makes perfect predictions from a diarization perspective.

\begin{figure}[t]
\subfloat[With collar tolerance\label{fig:eval_collar}]{\includegraphics[width=0.32\linewidth]{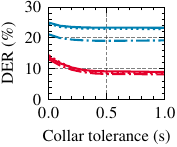}}\hfill
\subfloat[With closing applied to forced-aligned labels\label{fig:eval_closing}]{\includegraphics[width=0.32\linewidth]{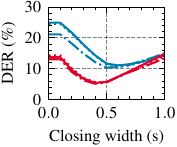}}\hfill
\raisebox{3pt}{\includegraphics[width=0.32\linewidth]{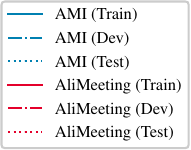}}
\caption{DERs (\%) computed between the original labels as ground truth and the forced-aligned labels as prediction.}
\label{fig:eval_tolerance}
\end{figure}

\begin{table*}[t]
\centering
\sisetup{detect-weight,mode=text}
\renewrobustcmd{\bfseries}{\fontseries{b}\selectfont}
\renewrobustcmd{\boldmath}{}
\newrobustcmd{\B}{\bfseries}
\setlength{\tabcolsep}{4pt}
\caption{DERs (\si{\percent}) on in-domain datasets. The cells with mismatched labeling criteria between training and evaluation are \colorbox{gray!20}{shaded}.}
\label{tbl:results_in_domain}
\vspace{-5pt}
\subfloat[SincNet]{%
\begin{tabular}{@{}lccccccccc@{}}
\toprule
&&\multicolumn{3}{c}{Labels for evaluation: ASR-oriented} & \multicolumn{5}{c}{Labels for evaluation: DIA-oriented}\\\cmidrule(l{\tabcolsep}r{\tabcolsep}){3-5}\cmidrule(l{\tabcolsep}){6-10}
& & AMI-MHM & AMI-SDM & AliMeeting & AMI-MHM & AMI-SDM & AliMeeting & MSDWild & VoxConverse \\
Trainind data & Closing & (Original) & (Original) & (Original) & (Forced-aligned) & (Forced-aligned) & (Forced-aligned) & (Original) & (Original) \\\midrule
\loosedata & & \B 19.95 & \B 24.33& \B 22.28 & \cellcolor{gray!20}{35.69}& \cellcolor{gray!20}{37.98}& \cellcolor{gray!20}{29.99} & 29.99 & \B 12.91\\\midrule
\tightdata & & \cellcolor{gray!20}{31.46} & \cellcolor{gray!20}{36.91} & \cellcolor{gray!20}{28.35} & \B 18.41 & \B 25.07 & \B 23.51 & \B 29.13 & 13.31\\
&$\checkmark$ & 20.64 & 27.23 & 23.59 & 36.09 & 42.18 & 32.94 & 32.38 & 13.85\\
\bottomrule
\end{tabular}%
}\\
\subfloat[ReDimNet-B2]{%
\begin{tabular}{@{}lccccccccc@{}}
\toprule
&&\multicolumn{3}{c}{Labels for evaluation: ASR-oriented} & \multicolumn{5}{c}{Labels for evaluation: DIA-oriented}\\\cmidrule(lr){3-5}\cmidrule(l){6-10}
& & AMI-MHM & AMI-SDM & AliMeeting & AMI-MHM & AMI-SDM & AliMeeting & MSDWild & VoxConverse \\
Trainind data & Closing & (Original) & (Original) & (Original) & (Forced-aligned) & (Forced-aligned) & (Forced-aligned) & (Original) & (Original) \\\midrule
\loosedata & & \B 16.65& \B 21.56 & \B 19.78 & \cellcolor{gray!20}{34.67}& \cellcolor{gray!20}{38.87}& \cellcolor{gray!20}{27.67}& 27.40& \B 11.83\\\midrule
\tightdata & & \cellcolor{gray!20}{28.88} & \cellcolor{gray!20}{32.47}& \cellcolor{gray!20}{26.29} & \B 15.03& \B 19.59 & \B 21.21 & \B 26.98 & 12.60\\
&$\checkmark$ & 17.14 & 21.62 & 21.47 & 34.01 & 37.82 & 31.11 & 30.84 &13.38\\
\bottomrule
\end{tabular}%
}
\end{table*}

As shown in \autoref{tbl:label_difference}, a substantial difference in DER arises solely from differences in annotation criteria: more than \SI{20}{\percent} for AMI and about \SI{14}{\percent} for AliMeeting.
A significant portion of these DERs can be attributed to missed speech, which often results from forced-alignment procedures that remove pauses within speech or trim silence buffers at the beginning and end of segments (see \autoref{fig:label_sample}).
This is quite striking and far from negligible, considering the DER values reported in recent state-of-the-art studies.
For example, DERs reported in pyannote.audio 3.1~\cite{plaquet2023powerset} are \SI{18.8}{\percent} for AMI-MHM, \SI{22.4}{\percent} for AMI-SDM, and \SI{24.4}{\percent} for AliMeeting.
As a result, when a diarization system successfully detects these pauses or silences, it may ironically be penalized in the evaluation, leading to inflated missed speech.
This indicates that missed speech due to the issues above could account for the majority of the total DER.
This observation raises serious concerns about the validity of evaluations conducted using ASR-oriented datasets, implying that they may not accurately reflect true diarization performance.

A common technique for mitigating label ambiguity in diarization evaluation is to apply collar tolerance when computing DER, which excludes the intervals around the start and end of each labeled segment from evaluation~\cite{nist_rt09}.
However, this approach does not mitigate issues arising from the presence or absence of pauses within reference segments.
\autoref{fig:eval_collar} shows how DER decreases as the forgiveness collar increases.
It reveals that collar tolerance is largely ineffective in compensating for the ambiguity present in ASR-oriented datasets.
This indicates that most of the labeling errors stem not from boundary misalignments, but from pauses within segments.
On the other hand, applying a closing operation to fill such pauses can recover roughly half of the gap, as shown in \autoref{fig:eval_closing}.
This suggests that this post-processing technique may be a useful heuristic when feeding ASR systems with output generated by models trained on DIA-oriented datasets.
Based on the analysis, we set the closing width to \SI{0.5}{\second} for AMI and \SI{0.6}{\second} for the others, unless otherwise stated.

\subsection{Impact of labeling strictness on model training}\label{sec:result_training}

\subsubsection{Evaluation on in-domain datasets}
\autoref{tbl:results_in_domain} shows the DERs on the in-domain datasets.
Both model architectures for AMI and AliMeeting achieved significantly better DERs when evaluated on test sets that followed the same annotation criteria as their respective training sets (cf. the bolded/shaded numbers).
This indicates that models tend to learn to output labels that align with the annotation policy used during training.

In contrast, for MSDWild and VoxConverse, where models were trained with DIA-oriented labels, the differences in DERs resulting from the evaluation criteria of AMI and AliMeeting were comparatively smaller.
This indicates that, despite inconsistencies in annotation policies across datasets, models implicitly learn dataset-specific conventions, such as whether to include pauses within speech segments.
While this may not pose a problem for in-domain evaluations, where training and test samples come from the same dataset as is often the case in recent neural diarization research, it presents a practical concern in real-world applications.
Specifically, if the model's outputs cannot be reliably classified as either ASR-oriented or DIA-oriented, it becomes difficult to determine how to post-process or interpret them for downstream tasks.
This unpredictability undermines usability and complicates integration into larger systems.

\subsubsection{Evaluation on out-of-domain datasets}
\begin{table*}[t]
\begin{minipage}{0.56\linewidth}
\centering
\sisetup{detect-weight,mode=text}
\renewrobustcmd{\bfseries}{\fontseries{b}\selectfont}
\renewrobustcmd{\boldmath}{}
\newrobustcmd{\B}{\bfseries}
\setlength{\tabcolsep}{4pt}
\caption{DERs (\si{\percent}) on out-of-domain datasets.}
\label{tbl:results_out_of_domain}
\vspace{-5pt}
\subfloat[SincNet]{%
\begin{tabular}{@{}lccccc@{}}
\toprule
&&\multicolumn{3}{c}{Labels for evaluation: ASR-oriented} & DIA-oriented\\\cmidrule(l{\tabcolsep}r{\tabcolsep}){3-5}\cmidrule(l{\tabcolsep}){6-6}
Training data & Closing& AISHELL-4 & DiPCo & NOTSOFAR-1 & DIHARD-3\\\midrule
\loosedata && \B 15.64 & \B 38.33 & 35.70 & 32.29\\\midrule
\tightdata && 27.04 & 43.56& \B 33.91 & \B 26.73\\
&$\checkmark$ & 16.03 & 38.53 & 37.71 & 35.27\\
\bottomrule
\end{tabular}%
}\\
\subfloat[ReDimNet-B2]{%
\begin{tabular}{@{}lccccc@{}}
\toprule
&&\multicolumn{3}{c}{Labels for evaluation: ASR-oriented} & DIA-oriented\\\cmidrule(l{\tabcolsep}r{\tabcolsep}){3-5}\cmidrule(l{\tabcolsep}){6-6}
Training data &Closing& AISHELL-4 & DiPCo & NOTSOFAR-1 & DIHARD-3 \\\midrule
\loosedata && 14.85 & 39.54 & \B 35.03 & 30.32\\\midrule
\tightdata && 26.26 & 43.05 & 36.81 & \B 25.50\\
&$\checkmark$ & \B 14.51 & \B 38.89 & 40.48 & 33.22\\
\bottomrule
\end{tabular}%
}
\end{minipage}\nextfloat\hfill
\begin{minipage}{0.4\linewidth}
\caption{Closing width $\delta$ (\si{\second}) that minimizes DER between output from the model trained using \loosedata and that using \text{Compound-tight}.}
\label{tbl:dataset_wise}
\centering
\begin{tabular}{@{}llcc@{}}
\toprule
&&\multicolumn{2}{c@{}}{Model architecture}\\\cmidrule(l){3-4}
Original labels&Dataset&SincNet&ReDimNet-B2\\\midrule
\multicolumn{4}{@{}l}{\textbf{In-domain datasets}}\\
ASR-oriented
&AMI-MHM & 0.56 & 0.56\\
&AMI-SDM & 0.55 & 0.58\\
&AliMeeting & 0.33 & 0.32\\
DIA-oriented
&MSDWild & 0.18 & 0.00\\
&VoxConverse & 0.22 & 0.00\\\midrule
\multicolumn{4}{@{}l}{\textbf{Out-of-domain datasets}}\\
ASR-oriented&AISHELL-4 & 0.46 & 0.46\\
&DiPCo & 0.35 & 0.21\\
&NOTSOFAR-1 & 0.15 & 0.00\\
DIA-oriented&DIHARD-3 & 0.39 & 0.31\\
\bottomrule
\end{tabular}
\end{minipage}
\end{table*}

\autoref{tbl:results_out_of_domain} shows the evaluation results on the out-of-domain datasets. 
As with the in-domain experiments, models trained on ASR-oriented labels achieved better DERs on AISHELL-4 and DiPCo, each of which is labeled according to ASR-oriented criteria.
However, in contrast to the in-domain case, for the dataset annotated with DIA-oriented labels, i.e., DIHARD-3, models trained on \tightdata achieved better DERs.
This indicates that models trained on \loosedata, including pause-inclusive labels, were unable to infer the annotation style of these new datasets and therefore failed to produce sufficiently tight segment boundaries.
In other words, the lack of alignment between the training and evaluation label styles led to suboptimal performance when the model encountered DIA-oriented data without having learned the appropriate segmentation behavior.
Note that although NOTSOFAR-1 is an ASR-oriented dataset, its results show inconsistencies between SincNet and RedimNet-B2, and it exhibits behavior characteristic of models trained with DIA-oriented data, such as DER degradation with closing.
One reason might be that the labels annotated for NOTSOFAR-1 are relatively tight, since they are based on ASR output, as shown in \autoref{tbl:dataset}.

\subsubsection{Effect of closing as a post-processing step}
As shown in \autoref{tbl:results_in_domain}, we observed that applying closing to the output of the model trained using \tightdata brought the DERs closer to those of the model trained using \loosedata.
This result is consistent with the findings in \autoref{sec:result_forced_alignment}.
It is important to note, however, that simple post-processing cannot achieve the reverse transformation from the output of the model trained on \loosedata to that of the model trained on \tightdata.
In contrast, for MSDWild and VoxConverse, the DERs for both models are already similar even without post-processing, and applying closing actually led to a degradation in performance.
This also implies that the models have already learned to produce DIA-oriented labels for these datasets, i.e., the models trained on \loosedata have learned a dataset-specific pause-filling strategy.
We discuss this phenomenon in detail in the following subsection.

\subsection{Dataset-wise behavior on pause filling}\label{sec:dynamics}
In this section, we further investigate the implicitly acquired pause-filling behavior by directly comparing the output from the models trained on \loosedata and \tightdata using the optimal closing width $\delta$ introduced in \autoref{sec:closing}.
A larger value of $\delta$ indicates a larger discrepancy between the outputs from the two models, i.e., the model trained using \loosedata tends to fill in longer pauses during inference.

As shown in \autoref{tbl:dataset_wise}, for in-domain datasets, the datasets annotated with ASR-oriented labels yielded dataset-specific optimal values of $\delta>0$.
In contrast, those with DIA-oriented labels resulted in smaller $\delta$ values, often dropping to zero with the high-performing architecture, i.e., ReDimNet-B2.
This clearly indicates that the models have effectively memorized the labeling guidelines of each dataset.

In contrast, for out-of-domain datasets, the labeling style (ASR-oriented or DIA-oriented) does not correlate meaningfully with the optimal $\delta$.
This confirms that the models are unable to determine, in out-of-domain datasets, how long a pause should be treated as part of a speech interval.
Considering this fact and the discussion regarding \autoref{tbl:label_difference} in \autoref{sec:result_forced_alignment}, this mismatch can significantly affect DER, especially under out-of-domain conditions.

These findings indicate that DER under such conditions may not reflect true diarization performance; at the very least, comparing absolute DERs between in-domain and out-of-domain settings is often meaningless, as the difference may simply reflect whether the model has aligned with the labeling style.
Therefore, it is essential that both training and evaluation be conducted using labels that follow a consistent, DIA-oriented annotation policy.

\subsection{From the perspective of streaming inference}
\begin{table*}
\centering
\setlength{\tabcolsep}{4pt}
\caption{DERs (\%) obtained by simulating streaming inference using only the final one second of each sliding window. The values in parentheses denote the change in DER relative to the offline results reported in \autoref{tbl:results_in_domain}.}
\label{tbl:streaming}
\begin{tabular}{@{}lcccccccc@{}}
\toprule
&\multicolumn{3}{c}{Labels for evaluation: ASR-oriented} & \multicolumn{5}{c}{Labels for evaluation: DIA-oriented}\\\cmidrule(lr){2-4}\cmidrule(l){5-9}
& AMI-MHM & AMI-SDM & AliMeeting & AMI-MHM & AMI-SDM & AliMeeting & MSDWild & VoxConverse \\
Trainind data & (Original) & (Original) & (Original) & (Forced-aligned) & (Forced-aligned) & (Forced-aligned) & (Original) & (Original) \\\midrule
\loosedata & 21.27 (+4.62) & 26.63 (+5.07) & 25.32 (+5.54) & 38.20 (+3.53) & 43.04 (+4.17) & 33.08 (+5.41) & 28.61 (+1.21) & 14.31 (+2.48)\\
\tightdata & 31.25 (\textbf{+2.37}) & 35.65 (\textbf{+3.18}) & 31.04 (\textbf{+4.75}) & 17.92 (\textbf{+2.89}) & 23.34 (\textbf{+3.75}) & 26.41 (\textbf{+5.20}) & 27.98 (\textbf{+1.00}) & 14.83 (\textbf{+2.23})\\
\bottomrule
\end{tabular}
\end{table*}

\begin{table}[t]
    \caption{ASR results in tcpWERs and ORC-WERs (\si{\percent}).}\label{tbl:results_asr}
    \vspace{-5pt}
    \centering
    \sisetup{detect-weight,mode=text}
    \renewrobustcmd{\bfseries}{\fontseries{b}\selectfont}
    \renewrobustcmd{\boldmath}{}
    \newrobustcmd{\B}{\bfseries}
    \subfloat[Multi-channel ASR with GSS and Whisper large-v3\label{tbl:results_asr_multi}]{%
    \begin{tabular}{@{}lcccc@{}}
    \toprule
    &\multicolumn{2}{c}{Closing}&\multicolumn{2}{c}{AMI-MDM}\\\cmidrule(l{\tabcolsep}r{\tabcolsep}){2-3}\cmidrule(l{\tabcolsep}){4-5}
    Training data& Mask & Beamforming & tcpWER & ORC-WER\\\midrule
    \loosedata &&& 30.93 & 23.84 \\\midrule
    \tightdata &&& 31.33 & 26.03 \\
    & $\checkmark$ & $\checkmark$ & 28.92 & 22.88 \\
    & & $\checkmark$ & \B 28.27 & \B 22.52 \\
    \bottomrule
    \end{tabular}%
    }\\
    \setlength{\tabcolsep}{3pt}
    \subfloat[Single-channel ASR with DiCoW\label{tbl:results_asr_single}]{%
    \begin{tabular}{@{}lccccc@{}}
    \toprule
    && \multicolumn{2}{c}{AMI-MHM} & \multicolumn{2}{c}{AMI-SDM}\\\cmidrule(l{\tabcolsep}r{\tabcolsep}){3-4}\cmidrule(l{\tabcolsep}){5-6}
    Training data& Closing & tcpWER & ORC-WER & tcpWER & ORC-WER\\\midrule
    \loosedata& & 21.72 & \B 15.69 & 31.61 & 20.58\\\midrule
    \tightdata& & 20.90 & 16.22 & 30.03 & 20.99\\
    &$\checkmark$ & \B 20.31 & \B 15.69 & \B 28.98 & \B 19.99\\
    \bottomrule
    \end{tabular}%
}
\end{table}

Including ASR-oriented labels in the training data requires the model to learn functionality similar to closing, which encourages it to rely on longer temporal context when making speech/non-speech decisions.
This becomes problematic in streaming applications, where prompt detection of speech onsets and offsets is crucial.
In this section, we investigate how different labeling strategies affect model performance under streaming conditions.

In the offline setting, predictions are aggregated over overlapping sliding windows of 10-second width with a one-second shift; thus, the final result for each moment is computed as the average over 10 evaluations.
In contrast, the streaming setting outputs only the final one second of each local window, in a manner similar to that of previous studies~\cite{coria2021overlapaware,rahou2024multi}.
To isolate the effect of streaming output constraints, clustering is performed in an offline manner in both cases, so the only difference lies in which part of the local diarization output is used to construct the global result.

\autoref{tbl:streaming} shows the results, with the values in parentheses indicating the performance degradation compared to the offline baseline in \autoref{tbl:results_in_domain}.
In all cases, models trained with DIA-oriented labels show smaller degradation.
This indicates that models trained on ASR-oriented labels, where speech segments often include pauses, tend to rely on future context to bridge silent gaps.
As a result, their performance degrades when only the final portion of a local window is used, as future context is not available in a streaming scenario.
When evaluated with ASR-oriented labels, models trained on \loosedata showed better absolute DERs due to consistency with the labeling criteria.
However, the degradation from offline to streaming inference is consistently smaller for models trained on \tightdata.
These results imply that training with DIA-oriented labels is preferable from the perspective of streaming diarization.

\subsection{From the perspective of ASR}
Some may be concerned that tight segment boundaries could negatively impact ASR performance.
We evaluate the effect of differences in the diarization results of the ReDimNet-based models on multi-channel and single-channel ASR.
We report the time-constrained concatenated minimum-perturbation word error rate (tcpWER)~\cite{von2023meeteval} and optimal reference combination word error rate (ORC-WER)~\cite{sklyar2022multi}.

\subsubsection{Multi-channel ASR}
For multi-channel ASR, we first applied guided source separation (GSS)~\cite{boeddeker2018front} to enhance each speaker's utterances using multiple distant microphones (AMI-MDM) based on the diarization results obtained using AMI-SDM.
For the model trained using \loosedata, the diarization results were used as is for GSS.
For the case of \tightdata, we use the diarization results under the following conditions: i) using the results as is for GSS, ii) applying closing before GSS, iii) using the results without closing for mask estimation in GSS, while applying beamforming based on the results with closing.
The third approach is designed to estimate time-frequency masks using purer speech/non-speech boundaries like in~\cite{wang2023ustc}, while beamforming is applied to segments after closing to prevent over-segmentation that may harm ASR.
We used the GPU-accelerated implementation~\cite{raj2023gpu} for GSS, and the Whisper large-v3 model~\cite{radford2023robust} to transcribe each enhanced utterance.

As shown in \autoref{tbl:results_asr_multi}, using diarization outputs from the models trained with \tightdata resulted in degraded ASR performance.
However, this issue was resolved by applying closing, which led to even better performance than that of the model trained with \loosedata.
Additionally, further improvement was observed when strictly estimated segments without closing were used for mask estimation, while segments after closing were used for beamforming and subsequent ASR to prevent over-segmentation.
These results indicate that, from a source separation perspective, predicting tight segment boundaries is advantageous.
Since closing can easily relax these boundaries when needed, there is no negative impact on ASR.
Therefore, it is preferable to train diarization models using datasets with tight boundary annotations.

\subsubsection{Single-channel ASR}
For single-channel ASR, we used the open-sourced diarization-conditioned Whisper (DiCoW)~\cite{polok2026dicow}\footnote{\url{https://huggingface.co/BUT-FIT/DiCoW_v1}}.
It takes a single-channel session recording and the corresponding diarization results as input.
In our experiment, we considered AMI-MHM and AMI-SDM as input recordings, 
The following three types of diarization results were evaluated in this study: the results from the model trained using \loosedata, those from the model trained using \tightdata with and without closing.

The results are shown in \autoref{tbl:results_asr_single}.
Unlike the cascade approach of multi-channel ASR, which performs separation followed by recognition, DiCoW takes the entire session audio as input.
As a result, the negative impact of over-segmentation in diarization is less severe compared to multi-channel ASR, and tcpWER even improved when using outputs from the model trained with \tightdata rather than those from the model trained with \loosedata (e.g., \SI{20.90}{\percent} vs. \SI{21.71}{\percent} for AMI-MHM).
Ultimately, using diarization results from the model trained on \tightdata with closing applied performed best in both datasets and evaluation metrics.
These findings indicate that even without source separation, training diarization models to predict tight segment boundaries is advantageous.

\section{Conclusion}
In this paper, we conducted a comprehensive investigation into the impact of using ASR-oriented datasets in speaker diarization research.
We showed that less strict segment boundaries in such datasets significantly affect evaluation metrics.
Furthermore, we demonstrated the undesirable behavior of models trained on loosely labeled data, including dataset-wise memorization of labeling strictness and degraded performance in streaming inference.
Finally, we showed that the potential negative impact of over-segmented labeling on ASR can be fully mitigated by applying closing as a post-processing step.

\clearpage
\IEEEtriggeratref{51}
\bibliographystyle{IEEEbib}
\bibliography{mybib,mfa}

\end{document}